\documentclass[numberedappendix]{emulateapj}

\usepackage{graphicx}
\usepackage{txfonts}

\slugcomment{Submitted to the Astrophysical Journal}

\shorttitle{Small-scale emergence of magnetic flux into the quiet atmosphere}
\shortauthors{Mart\' inez Gonz\'alez \& Bellot Rubio}

\begin{document}

\title{Emergence of small-scale magnetic loops through the quiet solar atmosphere}

\author{M.\ J.\ Mart\' inez Gonz\'alez\footnote{Formerly at Observatoire de Paris}}
\affil{Instituto de Astrof\'{\i}sica de Canarias, C/V\'{\i}a L\'actea s/n, 38200 La Laguna, Tenerife, Spain}
\email{marian@iac.es}
 
\and
 
\author{L.\ R.\ Bellot Rubio}
\affil{Instituto de Astrof\' isica de Andaluc\' ia (CSIC), Apdo.\ 
3004, 18080, Granada, Spain}

\begin{abstract}
We investigate the emergence of magnetic flux in the quiet Sun at very
small spatial scales, focusing on the magnetic connection between the
photosphere and chromosphere. The observational data consist of
spectropolarimetric measurements and filtergrams taken with the Hinode
satellite and the Dutch Open Telescope. We find that a significant
fraction of the magnetic flux present in internetwork regions appears
in the form of $\Omega$-shaped loops. The emergence rate is 0.02 loops
per hour and arcsec$^{-2}$, which brings $1.1 x 10^{12}$ Mx s$^{-1}$
arcsec$^{-2}$ of new flux to the solar surface. Initially, the loops
are observed as small patches of linear polarization above a granular
cell. Shortly afterwards, two footpoints of opposite polarity become
visible in circular polarization within or at the edges of the granule
and start to move toward the adjacent intergranular space. The
orientation of the footpoints does not seem to obey Hale's polarity
rules. The loops are continuously buffeted by convective motions, but
they always retain a high degree of coherence. Interestingly, 23\% of
the loops that emerge in the photosphere reach the chromosphere (16
cases out of 69). They are first detected in \ion{Fe}{1} 630~nm
magnetograms and 5 minutes later in \ion{Mg}{1} b 517.3~nm
magnetograms. After about 8 minutes, some of them are also observed in
\ion{Ca}{2} H line-core images, where the footpoints produce small
brightness enhancements.
\end{abstract}


\section{Introduction}

It is believed that solar magnetic fields are created in the
tachocline, the interface between the convection zone and the
radiative interior. Due to buoyancy instabilities, they move upward
and emerge into the solar atmosphere in the form of $\Omega$-shaped
flux tubes \citep{zwaan_85}. The largest emerging active
regions produce sunspots with magnetic fluxes in excess of $10^{21}$
Mx and lifetimes of several weeks to months. Smaller active regions
consist of pores and contain an order of magnitude less flux,
persisting over days to weeks. The smallest emerging regions detected
to date are the so-called ephemeral regions. They have fluxes between
$10^{18}$ and $10^{20}$ Mx and lifetimes in the range from hours to
days \citep[e. g.][]{martin_88, martin_79, hagenaar_03}.

Outside of active regions, the quiet Sun has proved to be full of
magnetic fields with strengths roughly in equipartition with the
photospheric convective flows \citep{lites_02, khomenko_03, ita_06,
david_07, marian_08}.  An important question is the origin of these
fields. \cite{lites_96} suggested that horizontal internetwork fields
represent concentrated loops of flux carried to the surface by the
upflows of granular convection or by magnetic buoyancy. In a recent
paper, \cite{marian_07} indirectly traced the emergence of magnetic flux 
and reconstructed, for the first time, the three dimensional topology 
of the magnetic field vector in quiet regions of the solar
photosphere. It was found that at least 20\% of the magnetic flux in
the quiet Sun is connected by low-lying magnetic loops. Later,
\cite{rebe_07} studied time series of spectropolarimetric observations
taken with the Solar Optical Telescope aboard {\it Hinode}. These
authors followed the time evolution of one magnetic loop in the
internetwork, showing that they appear on spatial scales smaller than
2$''$. \cite{ishikawa_07} and \cite{ishikawa+tsuneta2009} demonstrated 
that the emergence of magnetic flux on granular scales brings large 
amounts of horizontal fields to the photosphere both in plage regions 
and in the quiet Sun. Another form of flux emergence has been 
reported by \cite{david_08}. It involves the appearance and 
subsequent disappearance of what seem to be {\it vertical} fields 
at the center of granular cells.

The observations strongly suggest that a significant fraction of the
magnetic flux in the quiet Sun might be the result of the emergence of
small-scale magnetic loops. But, where do the loops come from? Are
they created by the global solar dynamo, by a local dynamo, or by
recycling of flux from decaying active regions? Is the emergence
process a local phenomenon confined to the photosphere or does the
magnetic flux reach higher atmospheric layers?

The answers to these questions bear important consequences for our
understanding of the magnetic and thermal structure of the solar
atmosphere. For example, \cite{javier_04} claim that the magnetic
energy stored in the quiet photosphere is sufficient to balance the
radiative losses of the chromosphere. Quiet Sun magnetic fields are
excellent candidates to solve the chromospheric and coronal heating
problem, but a mechanism capable of transferring their energy to the
upper layers has not been identified yet. From a theoretical point 
of view, it is not clear whether the fields of the quiet solar
photosphere can rise to the chromosphere. \cite{isobe_08} have
presented MHD simulations in which the magnetic field emerges into the
photosphere in the form of small-scale $\Omega$-loops. They reach the
chromosphere and get reconnected with the local expanding vertical
magnetic fields, heating the plasma and generating high frequency MHD
waves that propagate into the corona. However, the magnetoconvection
simulations of \cite{stein_06} show $\Omega$-loops that disintegrate 
as they rise through the solar atmosphere. 

These discrepancies emphasize the need for observational studies aimed
at determining whether magnetic fields emerging into the quiet
photosphere are able to reach higher atmospheric layers. Here we use
multi-wavelength observations taken by {\it Hinode} and the Dutch Open
Telescope to address this question. We also characterize the physical
properties of small-scale magnetic loops in the quiet Sun, providing
estimates of their magnetic fluxes, emergence rates, lifetimes, sizes,
and velocities.

\section{Observations and data reduction}

The data analyzed in this paper consist of time series of polarimetric
and imaging observations of quiet Sun regions at disk center. They were acquired in seven different days (25-29 September, 1 and 6
October 2007) using the instruments of the Solar Optical Telescope
aboard Hinode \citet{kosugi2007} and the Dutch Open Telescope
\citet[DOT;][]{rutten_etal04} at Observatorio de El Roque de Los
Muchachos (La Palma, Spain). The observations belong to the {\it
Hinode} Operation Plan 14, entitled ``Hinode/Canary Islands
campaign''.

The {\it Hinode} spectro-polarimeter \citep[SP;][]{litesetal_01}
recorded the full Stokes vector of the pair of \ion{Fe}{1} lines at
630~nm in a narrow field of view (FOV) of $2.7''\times 40.6''$. This
region was scanned with a cadence of 28~s during 2-6 hours per day (Table \ref{tabla_obs}). The exposure time per slit position was set to 1.6~s to
track very rapid events. However, this mode of operation also led to
a noise level of $1.7 \times 10^{-3}$ in units of the continuum
intensity $I_{\rm c}$. With a pixel size of $0.16''$ along the slit
and $0.15''$ perpendicular to it, the SP measurements have a spatial
resolution of about $0.32''$.

The Hinode Narrowband Filter Imager (NFI; Tsuneta et al. 2008)
acquired Stokes I and V filtergrams in the wings of the chromospheric
Mg I b 517.3 nm line, $\pm 11.5$~pm away from its center. The NFI was
operated in shutterless mode to reach an effective exposure time of
9.6 s per wavelength and polarization state, covering a FOV of 
$15.4\arcsec \times 65.3\arcsec$. The original filtergrams had a pixel size of 0.08$''$, but we
rebined them to the SP pixel size in order to further reduce the
noise. The {\it Hinode} Broadband Filter Imager
\citep[BFI;][]{tsunetaetal_08} acquired simultaneous images of the
photosphere in the CN bandhead at 388.3~nm (filter width of 0.52~nm)
and the chromosphere in the Ca\,{\sc ii}\,{\sc H} line at 396.85~nm
(filter width of 0.22~nm). The exposure times were 0.1 s and 0.3 s,
respectively. The BFI covered a region of $19.2'' \times 74.1''$ 
with a pixel size of $0.055''$. Both the NFI and the BFI took images with 
a cadence of 30 s.

The area scanned by the SP represents a small part of the total 
FOV of the NFI and the BFI. Therefore, we have cospatial and 
cotemporal observations of the quiet Sun tracing different 
heights in the atmosphere.

The DOT observed photospheric and chromospheric layers by means of a
tunable Lyot filter that scanned the intensity profile of the
H$\alpha$ line at five wavelength positions ($\pm 0.7$, $\pm 0.35$,
and 0~\AA\/). The passband of the filter was 0.25~\AA\/. Speckle
bursts of 100 frames were taken at each wavelength position every
30~s. Following the standard reduction procedure at the DOT, the
individual filtergrams were reconstructed using a speckle masking
technique \citep[see][for details]{rutten_etal04}. The reconstructed
images cover a FOV of $91.2'' \times 89''$ and have a spatial
resolution of about 0.2\arcsec. The DOT and {\it Hinode} carried out 
simultaneous observations, but there is little overlap between them because 
of bad weather conditions.

The SP data have been corrected for dark current, flat-field, and
instrumental cross-talk using the sp\_prep.pro routine included in the
SolarSoft package. The algorithm applied to the Hinode filtergrams
(fg\_prep.pro) removed dark current, hot pixels, and cosmic rays. The
spectropolarimetric maps and the various filtergrams have been aligned 
with pixel accuracy using the granulation, G-band bright points, and 
network elements as a reference.

\begin{figure*}[!t]
\centering{\includegraphics[width=0.9\textwidth,bb=51 192 527 662]{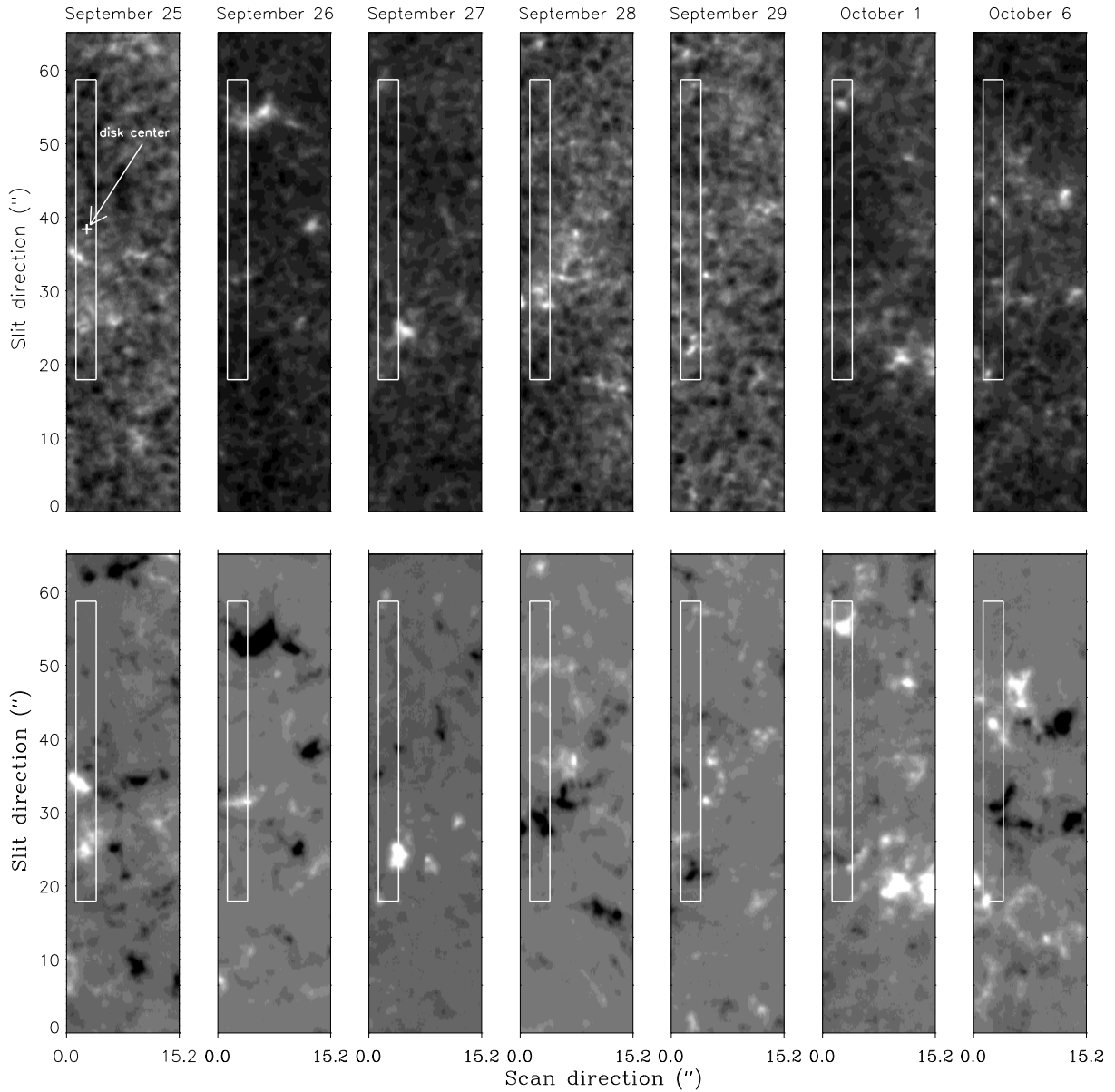}}
\caption{Quiet Sun regions observed during HOP 14. {\it Top:} \ion{Ca}{2}~H  
line core maps averaged over each day. {\it Bottom:} Same, for the \ion{Mg}{1}~b 
magnetograms. White and black represent positive and negative polarity 
magnetic fields. The white rectangles indicate the areas scanned with 
the spectro-polarimeter. North is up and West to the right.} \label{calcio}
\vspace*{-1em}
\end{figure*}

In Fig.\ \ref{calcio} we show time-averaged \ion{Ca}{2}~H filtergrams and \ion{Mg}{1}~b magnetograms
for the seven days of observation. The \ion{Ca}{2} H images have been trimmed to the size of the \ion{Mg}{1} b FOV. The rectangles represent the areas scanned by the
SP. Note the absence of strong brigthenings in the \ion{Ca}{2} H maps, 
as expected for very quiet regions largely devoid of network elements. 

\begin{figure*}[!ht]
\hspace{1.1cm}{\large$\Delta$ t (s) $\rightarrow$}\\
\includegraphics[width=\textwidth,bb= 28 19 492 120]{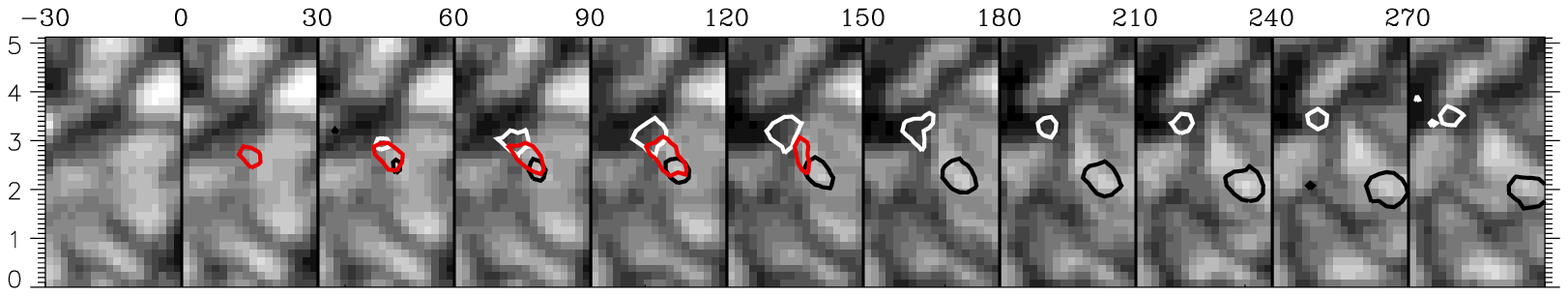}
\includegraphics[width=\textwidth,bb= 28 19 492 120]{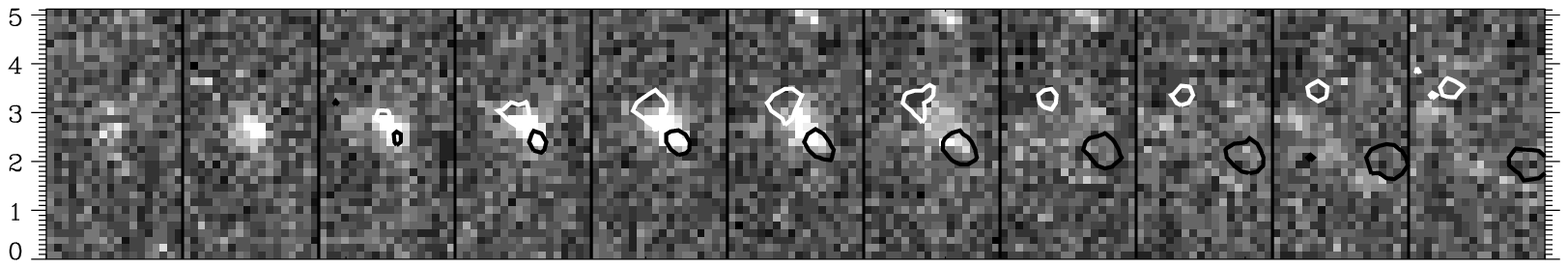}
\includegraphics[width=\textwidth,bb= 28 19 492 120]{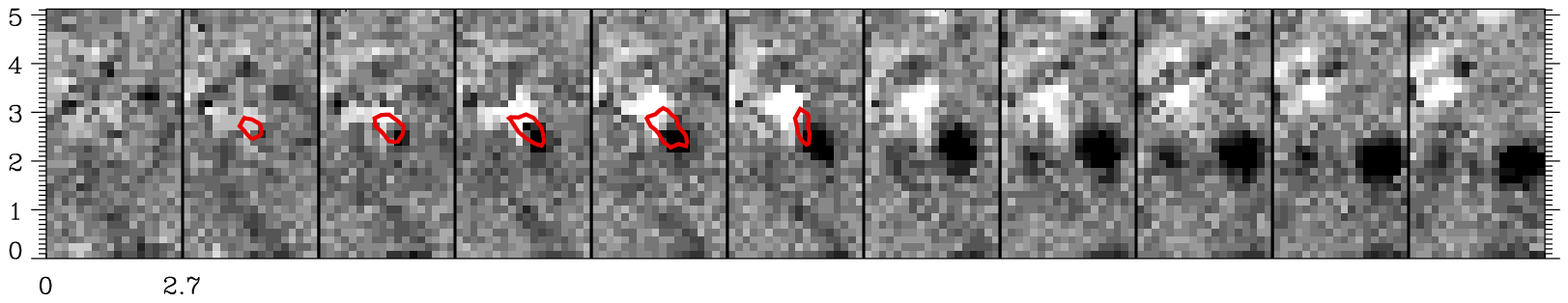}
\caption{Emergence of a small-scale magnetic loop in the quiet solar photosphere. 
Time runs from left to right. {\it Top:} Maps of continuum intensity
at 630~nm. {\it Middle:} total linear polarization in the 630.25 nm
line, saturated at 0.3~pm. {\it Bottom:} total circular polarization
in the 630.25 nm line. The signals are clipped at $\pm 0.1$~pm. Red
contours represent linear polarization larger than $0.22$~pm. Black
and white contours indicate circular polarization signals stronger
than $\pm 0.1$~pm. Both x and y-axis are in arcsec.}
\label{primer_ejemplo}
\end{figure*}

\begin{table}[!b]
\begin{center}
\caption{Log of the observations}
\label{tabla_obs}
\begin{tabular}{ccccc}
\tableline
Date & 1st period & 2nd period & Observed & Detected \\
 & (UT) & (UT) & time & loops\\
\tableline
25$^\mathrm{th}$ Sep & 13:00-15:59 & & 3.0 h & 18\\
26$^\mathrm{th}$ Sep & 08:15-14:14 & & 6.0 h & 16\\
27$^\mathrm{th}$ Sep & 06:16-09:59 & 11:25-13:59 & 6.3 h & 7\\
28$^\mathrm{th}$ Sep & 07:00-09:59 & 11:20-13:59 & 5.7 h & 11\\
29$^\mathrm{th}$ Sep & 06:51-09:44 & & 2.9 h & 5\\
1$^\mathrm{st}$ Oct & 08:21-10:09 &  & 1.8 h & 3\\
6$^\mathrm{th}$ Oct & 08:01-10:18 &  & 2.3 h & 9\\
\tableline
\end{tabular}
\end{center}
\end{table}

\section{Data analysis}

\subsection{Detection of magnetic loops in different layers}

Loop-like magnetic structures leave clear signatures in
spectropolarimetric maps: linear polarization flanked by two circular
polarization signals of opposite polarity. One of the main goals of
this paper is to trace the possible ascent of small-scale magnetic
loops through the solar atmosphere. To this end, we use photospheric
and chromospheric observables.

The SP data make it possible to investigate the topology
of the field in the photosphere. The information is
complemented by the CN filtergrams, where bright points associated
with magnetic fields are easily visible. We define the total circular
polarization as the integral of the unsigned \hbox{630.25 nm} Stokes
$V$ spectrum. The integration is carried out in the wavelength range
from $-17.25$~pm to $+30.31$~pm. The total linear polarization is 
computed as the integral of the Stokes $L = \sqrt{Q^2+U^2}$ profile of
\ion{Fe}{1}~603.25~nm, using the same initial and final wavelengths.

We visually inspect the polarization maps to search for weak linear
signals between two patches of circular polarization with opposite
polarity. Since in general the linear polarization signals are very
small in the quiet Sun, we ascribe those cases to loop-like structures
only after corroborating that the linear polarization is produced by
symmetric Stokes $Q$ and $U$ profiles.

Figure \ref{primer_ejemplo} shows the first stages of the emergence of
a small-scale magnetic loop as seen in the \ion{Fe}{1} 630.25~nm
line. The different panels represent continuum intensity (top), total
linear polarization (middle), and total circular polarization
(bottom).  Red contours indicate enhanced linear polarization.  Black
and white contours mark the location of strong negative and positive
circular signals. At $\Delta t=0$~s, a patch of linear polarization
shows up at the center of the image. It corresponds to the horizontal
part of a magnetic structure emerging into the photosphere. Between
$\Delta t=30$ and 60 s, two circular polarization patches of opposite
polarity appear next to it, at the edge of a granular cell. The fact
that the linear signal is detected earlier than the circular
polarization indicates that the magnetic structure has the shape of an
$\Omega$-loop: the apex creates linear polarization and the vertical
fields of the footpoints give rise to circular polarization \citep{rebe_07}. 
The linear polarization disappears below the noise at $\Delta t \sim
150$~s while the footpoints separate with time. This sequence of
events is consistent with a loop that emerges and travels up in the
atmosphere. The distance between the footpoints keeps increasing until
they reach the edges of the area scanned by the SP. The subsequent
evolution of this loop will be studied in \S \ref{description}.

The Stokes $I$ and $V$ filtergrams acquired in the red and blue wings
of the \ion{Mg}{1}~b line give information about the upper 
photosphere/temperature minimum region \citep{lites_88}. We have 
used them to construct longitudinal magnetograms ($M$) and Dopplergrams
($D$) as
\begin{eqnarray}
&M =& \frac{1}{2} \left( \frac{V_{\rm b}}{I_{\rm b}} - 
                         \frac{V_{\rm r}}{I_{\rm r}} \right) \\
&D =&\frac{I_{\rm r}-I_{\rm b}}{I_{\rm r}+I_{\rm b}} ,
\label{eq1}
\end{eqnarray}
where the subscripts r and b represent the measurements at $+11.5$ and
$-11.5$~pm from line center, respectively. To first order, the
magnetograms computed in this way are not affected by mass
motions. The quantities $M$ and $D$ have been transformed into
magnetic flux densities and line-of-sight velocities according to
\begin{eqnarray}
\phi &\approx& 9063 \, M, \\
v_{\rm LOS} &\approx& 10.48 \, D - 0.28, \label{dopplergram}
\end{eqnarray}
with $\phi$ in Mx~cm$^{-2}$ and $v_{\rm LOS}$ in km~s$^{-1}$. Equation
\ref{dopplergram} is valid in the range $-2 < v_{\rm LOS} 
< 2$~km~s$^{-1}$. These expresions have been obtained through calibration of the \ion{Mg}{1} b line shape in the Fourier Transform Spectrometer atlas of the quiet Sun \citep{ftsatlas} and give only rough estimates of the magnetic flux density and velocity at the height of formation of the \ion{Mg}{1} b measurements. Since Stokes Q and U were not recorded, the Mg I
magnetograms can only be used to detect relatively vertical fields such as those
expected at the footpoints of magnetic loops. By definition, fields
pointing towards the observer and upflows will both be positive. Note
that our sign convention for the velocity differs from that commonly
used in astrophysics.

In the chromosphere we do not have polarimetric information. However,
magnetic fields can be detected through brightness enhancements in the
\ion{Ca}{2} H filtergrams. The passband of the {\it Hinode}
\ion{Ca}{2} H filter includes a significant photospheric contribution,
but it has a long tail that extends well into the chromosphere
\citep{carlsson_etal07}. Finally, information on the upper
chromosphere is provided by the H$\alpha$ measurements taken at the
DOT. We have used them to construct Dopplergrams at different heights
in the chromosphere. 

\subsection{Photospheric magnetic flux}

We determine the magnetic flux density from the Stokes $V$ profiles of
the \ion{Fe}{1} 630~nm lines using the weak field approximation
\citep[e.g.,][]{landi92}
\begin{equation}
V(\lambda) = - \phi \, C \, \frac{\partial I(\lambda)}{\partial \lambda},
\end{equation}
where $\phi = f B \cos \gamma$ is the longitudinal flux density, $f$
the filling factor, $B$ the field strength, $\gamma$ the inclination
of the field with respect to the vertical, $C=4.6686\times
10^{-13}\lambda_0^2 \bar{g}$ a proportionality constant that depends
on the central wavelength $\lambda_0$ and the effective Land\'e factor
$\bar{g}$ of the transition, and $I$ represents the intensity
profile. The units of $\phi$ are Mx~cm$^{-2}$ when $\lambda_0$ is
expressed in \AA\/.

The longitudinal flux density is obtained from a least-squares 
minimization of the form
\begin{equation}
\frac{\partial}{\partial \phi}\left[\sum_i \left( V_i+\phi C\frac{\partial I}{\partial \lambda}_i\right)^2\right]=0 ,
\label{weakfield}
\end{equation}
which uses all the wavelength samples across the profile (index $i$) 
and is therefore more accurate that determinations based on single 
magnetogram measurements. This calculation is repeated for each 
pixel and each spectral line separately. The final result is 
\begin{equation}
\phi=-\frac{\sum_i \frac{\partial I}{\partial \lambda}_i V_i}{C\sum_i 
(\frac{\partial I}{\partial \lambda}_i)^2}
\label{eq2}
\end{equation}

To estimate the uncertainty caused by photon noise we
simulated a data set containing only gaussian noise with a standard
deviation of $1.7\times 10^{-3}$ $I_{\rm c}$. The analysis of this
data set using Eq.~\ref{eq2} leads to a gaussian-shaped histogram for
the magnetic flux density which is centered at 0 and has a standard
deviation of $\sigma_\phi =3.5$ Mx~cm$^{-2}$.

\begin{figure}[!t]
\centering{ \includegraphics[width=.96\columnwidth,bb=51 12 480 338]{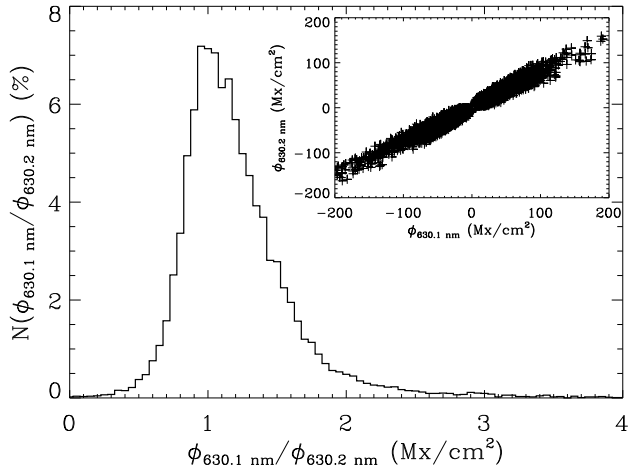}}
\caption{Histogram of the ratio between the magnetic flux density inferred from \ion{Fe}{1} 
630.15~nm and 630.25~nm. The inset shows a scatter plot of the values derived from the two spectral lines.}
\label{hist_ratio_flux}
\end{figure}

In Figure \ref{hist_ratio_flux} we check the assumption made on the
Zeeman regime. The plot shows an histogram of the ratio between the
magnetic flux densities derived from \ion{Fe}{1} 630.15 and 630.25~nm. Only the Stokes V spectra at the footpoints 
of the loops having amplitudes above 5 times the
noise level have been considered. As can be seen, the histogram is
narrow and peaks at 1. The figure also shows a scatter plot of the magnetic flux
densities obtained with the two lines. The fact that most of the
points are located near the diagonal implies that in the majority of 
cases both spectral lines measure the same magnetic flux. This 
strongly supports the idea that the fields are weak.

\begin{figure*}[!t]
\includegraphics[width=\textwidth, bb= 46 169 565 752]{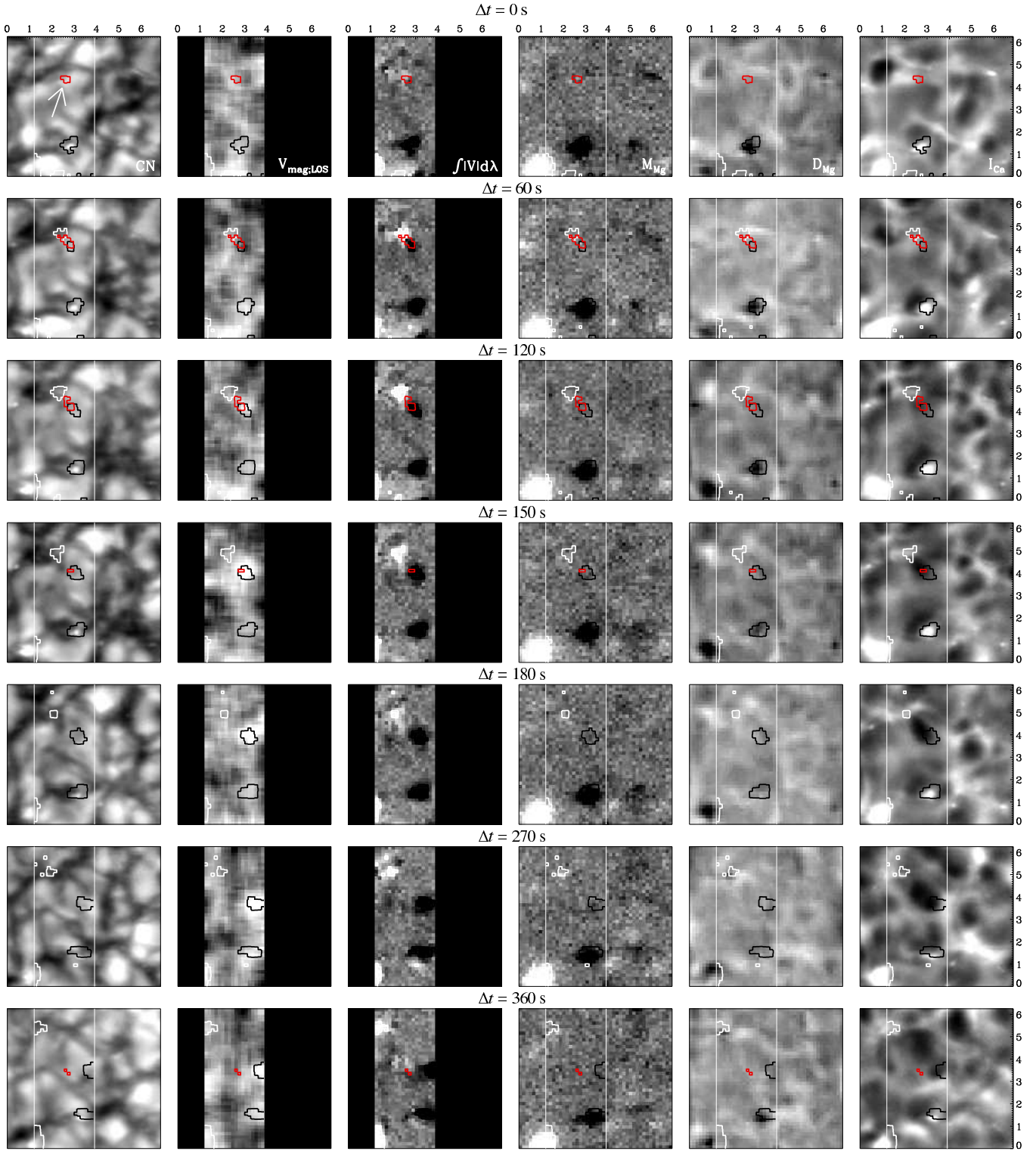}
\caption{Evolution of the small-scale magnetic loop ml09 that emerges in the 
photosphere and reaches the chromosphere. The panels show, from left
to right, CN intensity, \ion{Fe}{1} 630.25~nm Stokes $V$
zero-crossing velocity (scaled to $\pm 4$~km~s$^{-1}$, with upflows
being light), total circular polarization in \ion{Fe}{1} 630.25~nm
(clipped to $\pm 0.1$~pm), magnetic flux density from 
\ion{Mg}{1}~b 517.3~nm (scaled to $\pm 22.7$~Mx~cm$^{-2}$), LOS
velocity from the \ion{Mg}{1}~b Dopplergram (scaled between $-0.5$ and
$0$~km~s$^{-1}$, with stronger downflows appearing darker), and
\ion{Ca}{2} H line-core intensity.  The vertical white lines represent the
SP scan. The scales in the horizontal and vertical axes are arcsec. 
Red contours indicate areas with linear polarization signals 
in \ion{Fe}{1} 630.25~nm larger than 0.22~pm. Black and white contours
represent circular polarization signals in the photosphere stronger
than $\pm 0.1$~pm. Blue and turquoise contours show magnetic flux
densities from the \ion{Mg}{1}~b magnetograms larger than $\pm
13.6$~Mx~cm$^{-2}$. The x and y-axis are in arcsec.}
\label{ej1_crom}
\vspace*{-1em}
\end{figure*}

\begin{figure*}[!t]
\includegraphics[width=\textwidth, bb= 46 169 565 752]{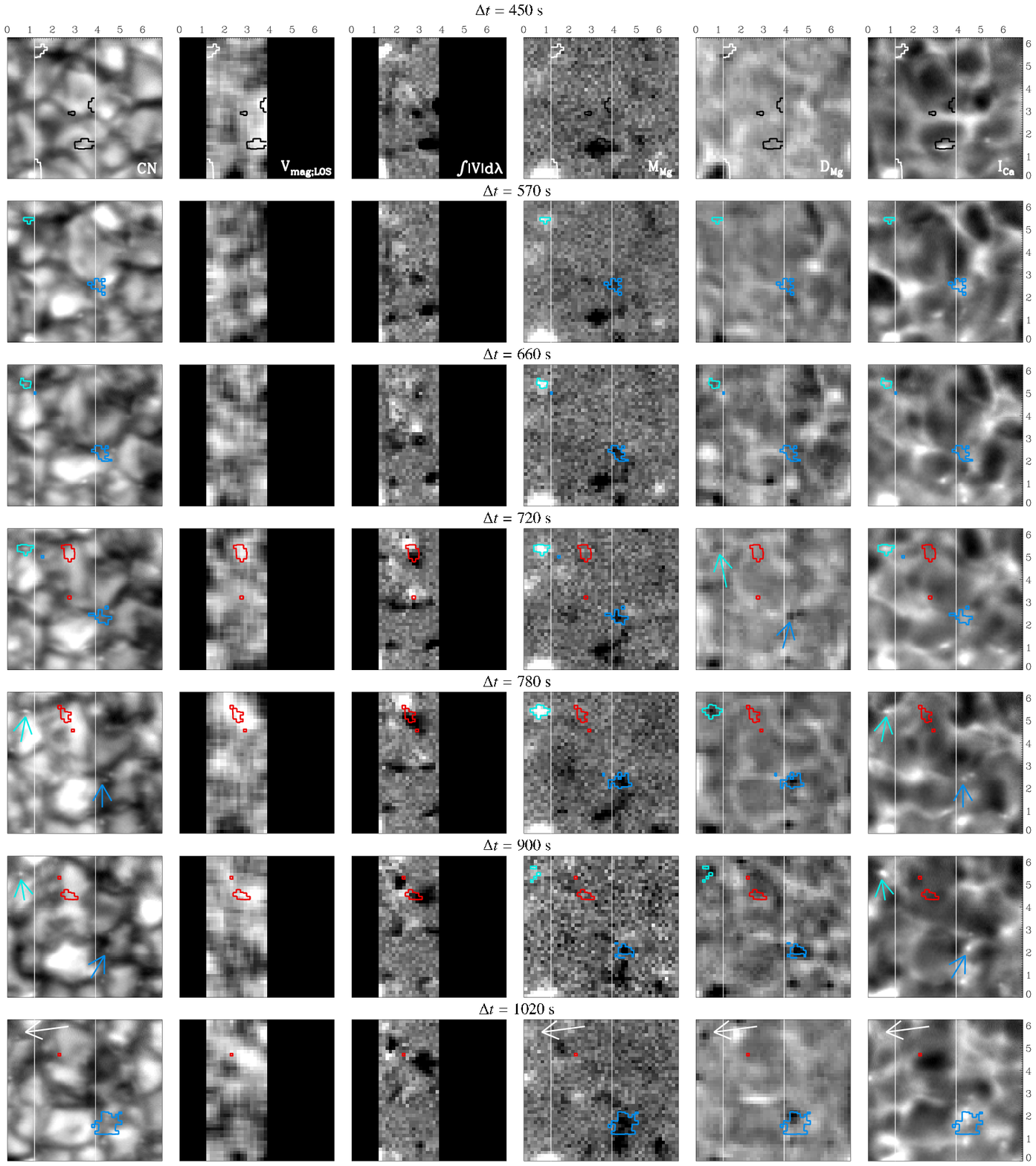}
\caption{Continuation of Fig.\ \ref{ej1_crom}. }
\label{ej1_crom2}
\vspace*{-1em}
\end{figure*}

\section{Emergence of small-scale magnetic loops in the quiet Sun}

During 28 hours of observations we have detected the appearance of 69
loop-like structures in the $2.7\arcsec \times 40.6\arcsec$ region
scanned with the Hinode spectropolarimeter. Table \ref{tabla} summarizes their properties. The observed loops followed very similar evolution patterns. In this section we
describe specific cases to illustrate the emergence process,
distinguishing between loops that rise to the chromosphere and loops
that remain low-lying.

\subsection{Magnetic loops emerging up to the chromosphere}
\label{description}

One of the clearest examples of a loop that reached the chromosphere
was loop ml09, observed on September 25, 2007 at 14:39 UT. Its evolution is
summarized in Figs.\ \ref{ej1_crom} and \ref{ej1_crom2}. Each row represents a time step,
for a total of 1020 s. From left to right we show CN filtergrams,
photospheric LOS velocities derived from the Stokes $V$ zero-crossing
shifts of \ion{Fe}{1} 630.25~nm, total circular polarization maps in
\ion{Fe}{1} 632.05~nm, magnetic flux densities computed from 
\ion{Mg}{1}~b 517.3~nm, LOS velocities at the height of formation of
the \ion{Mg}{1}~b line, and \ion{Ca}{2} H line-core filtergrams. 
Positive velocities indicate upflows. Note that the time steps are 
not evenly spaced; rather, we have adjusted them to better describe 
the various phases of the process.

The red contours mark regions of large linear polarization signals in
\ion{Fe}{1} 630.25 nm. When the footpoints of the loop are visible in
the photosphere, we plot contours of \ion{Fe}{1} 630.25~nm circular
polarization in black and white for the negative and positive signals,
respectively. When the footpoints are only visible in the
\ion{Mg}{1}~b magnetogram we plot them in blue and turquoise. 

The first row of Fig.\ \ref{ej1_crom} shows the emergence of linear
polarization above a granule (cf.\ the white arrow in the CN
filtergram). It is caused by the horizontal part of the loop reaching
the photosphere. The footpoints are not yet visible but will appear 
30~s later, very close to the patch of linear polarization. At $\Delta t=
60$ the loop is completely formed. It emerges in a granular region,
perhaps because the upward granular motions help the field lines to
rise from below the solar surface
\citep{mark_07}. The LOS velocity maps show photospheric upflows 
at the position of the footpoints, confirming that the loop is
rising. For the moment, however, the magnetic field remains in the
lower photosphere: we do not observe circular polarization signals in
the \ion{Mg}{1} b magnetograms or brightenings in Ca\,{\sc ii}\,{\sc
H} that could be associated with the loop. Interestingly, the strong
photospheric network element with negative polarity located towards
the bottom of the scan, at a height of about one fourth of the
displayed FOV, is well observed both in the \ion{Mg}{1} b magnetograms
and the Ca filtergrams, and will remain so during most of the time
sequence. This nicely illustrates the capabilities of our
observations: magnetic structures that are visible in the \ion{Fe}{1}
polarization maps but not in \ion{Mg}{1}~b or \ion{Ca}{2} H are
intrinsically lower in the solar atmosphere.

Between $\Delta t= 150$ s and $\Delta t=180$ s the linear polarization
signals disappear below the noise. The positive footpoint has drifted
to an intergranular lane and is concentrated, whereas the negative
footpoint continues to be rooted in the granule and is more
diffuse. The distance between them increases steadily. We still see
upward motions in the photospheric velocity maps. Since the loop is
moving to higher layers, it is reasonable to conclude that the linear
polarization disappears because the apex of the loop leaves the
formation region of the 630~nm lines. However, no traces of the 
loop are detected yet in \ion{Mg}{1}~b or \ion{Ca}{2} H.

At $\Delta t= 270$ s, weak circular polarization signals cospatial
with the photospheric footpoints are observed in the \ion{Mg}{1} b
magnetograms for the first time. This indicates that the loop has
reached the upper photospheric/lower chromospheric layers where the
central part of the \ion{Mg}{1}~b line forms. Interestingly, the
\ion{Mg}{1}~b Dopplergram exhibits downflows of about $-0.4$~km~s$^{-1}$ 
at the position of the positive footpoint. The downflows could
represent plasma moving along the legs of the loop as the whole
structure reaches high atmospheric layers. These motions may be
essential for the loop to get rid of part of its mass before it can
emerge into a less dense medium.

\begin{figure}
\centering{ \includegraphics[width=.97\columnwidth,bb=25 10 495 344]{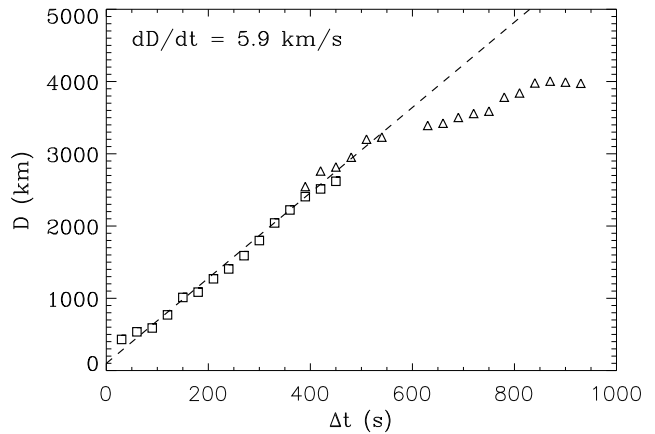}}
\caption{Footpoint separation as a function of time for the loop displayed in 
Figure \ref{ej1_crom}. The initial time corresponds to the appearence of
the linear polarization signal (the top of the loop). The distances
between the footpoints in the \ion{Fe}{1} 630~nm polarization maps 
and the \ion{Mg}{1}~b magnetograms are indicated with squares and
triangles, respectively. The linear fit corresponds to the first 
500~s of the loop evolution.}
\label{dist}
\end{figure}

At $\Delta t=570$ s, the loop is nearly out of the region scanned by
the SP. The signals in the \ion{Mg}{1}~b magnetogram are much more
intense and correspond to footpoints rooted in intergranular
lanes. From now on the distance between the footpoints will increase,
but at a slower rate than when they were crossing granular
structures. This inflection point can be seen in Fig.\ \ref{dist},
where we plot the footpoint separation as a function of time. The
distance is computed only when the two footpoints are visible, both in
the \hbox{630 nm} maps (squares) and in the \ion{Mg}{1}~b magnetograms
(triangles). In the first 500 s of the loop evolution, the distance
between the footpoints increases linearly at a rate of
5.9~km~s$^{-1}$. Therefore, the mean velocity of the footpoints is
2.95~km~s$^{-1}$, a value compatible with the motion of the granular
plasma. The linear increase of the separation with time is a
common feature of the loops and indicates that they do not undergo a
free random walk (otherwise the distance would increase as the square
root of time). Towards the end of the loop evolution the separation
rate slows down, coinciding with the arrival of the footpoints to
intergranular lanes. Summarizing, the loop emerges in a granule and
the horizontal granular motions drive the magnetic field lines to the
closest intergranular space, where strong downdrafts capture and
stabilize them. When this happens, the separation between the
footpoints is about 4000~km.

At $\Delta t=660$ s, the footpoints are clearly visible in the
\ion{Mg}{1}~b magnetogram and exhibit downflows in the \ion{Mg}{1} 
b Dopplergram. The whole structure is rising because the
footpoints continue to separate. However, no brightenings are detected
in the \ion{Ca}{2} H filtergrams. We mention in passing that a new
loop appear in the FOV at this time, very close to the site of
emergence of the structure we are describing in detail. They show
a linear polarization signal in between opposite polarities, which
makes it easy to identify.

At $\Delta t=780$ s, the loop has reached the chromosphere since we
observe two \ion{Ca}{2} H brightenings associated with the footpoints
(in Fig.\ \ref{ej1_crom2}, the contours have been substituted by arrows
for clarity). The legs of the loop still show downflows in the
\ion{Mg}{1}~b Dopplergrams and, for the first time, bright points
are observed in the CN images at the position of the footpoints.

The last panels of Fig.\ \ref{ej1_crom2} displays the beginning of the
loop decay. The positive footpoint is very weak, although it still
shows downflows at the height of formation of the \ion{Mg}{1}~b
measurements. It will disappear below the noise level, together with
the downdrafts, at the end of the sequence. The negative footpoint is
approaching a negative polarity patch with whom it will eventually
mix. The negative footpoint shows downflows and is associated with a
bright point in \ion{Ca}{2}~H. These features will survive the
disappearance of the footpoint thanks to the interaction with the
network element.

\subsection{Low-lying magnetic loops}

\begin{figure*}
\includegraphics[width=\textwidth,bb=49 437 560 768]{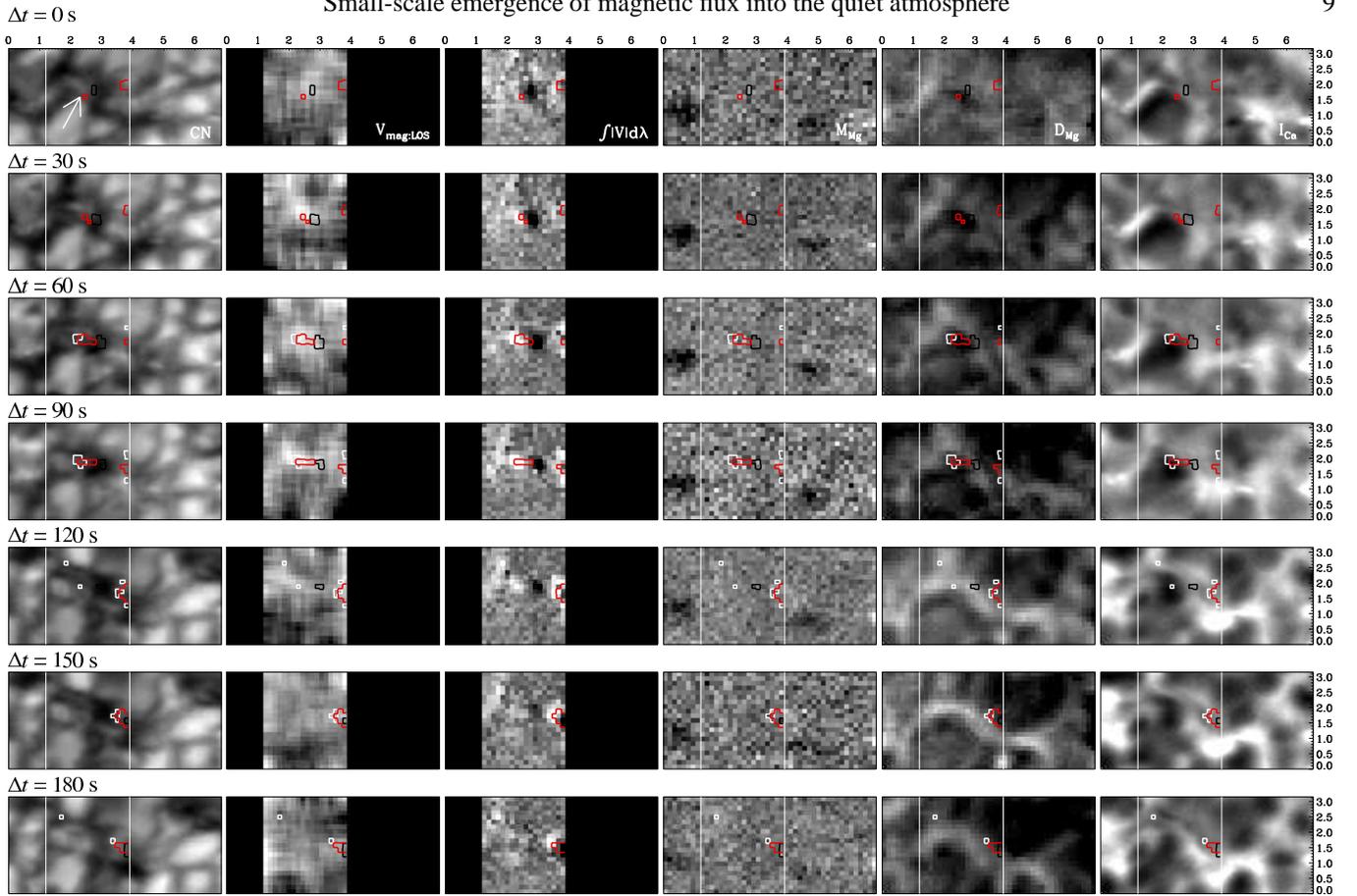}
\caption{Evolution of the small-scale magnetic loop ml23 that never reached 
chromospheric heights. The loop appeared on September 26, 2007 at 11:48 UT. 
The panels are arranged in the same way as in Fig.\ \ref{ej1_crom}.
\label{ej1_nocrom}}
\end{figure*}

In this section we present a typical example of a loop which do not show 
chromospheric signatures and thus remain low-lying. Figure
\ref{ej1_nocrom} shows all the data available for this loop (ml23), arranged
as in Fig.\ \ref{ej1_crom}.

In the first frame, a patch of linear polarization is observed to
emerge at the border of a granule (see the white arrow). The
footpoints can already be detected in the intergranular lane, but 
they are very weak. At this time the photospheric velocity map exhibits
a patch of upflows at the position of the loop, confirming its rise
through the solar atmosphere.

At $\Delta t=90$ s the loop reaches its largest extent (520 km) while
the upflows start to weaken. The linear polarization and the upward
plasma motions are almost gone by $\Delta t = 120$ s. In the next
frame, at $\Delta t = 150$ s, the loop is no longer seen. The
evolution of the loop is so rapid that it appears and disappears
almost at the same place. Interestingly, the footpoints never 
approach each other. This rules out submergence below the solar
surface as the cause of the loop disappearance.

All the loops that stay in the photosphere show very similar
behaviors. In general, the evolve very quickly, disappearing 
not far from the region where they emerged. None of these loops 
exhibit downflows in the \ion{Mg}{1}~b line or brightenings in 
the CN or Ca filtergrams.

\begin{figure*}[!ht]
\includegraphics[width=\textwidth,bb=60 121 560 765]{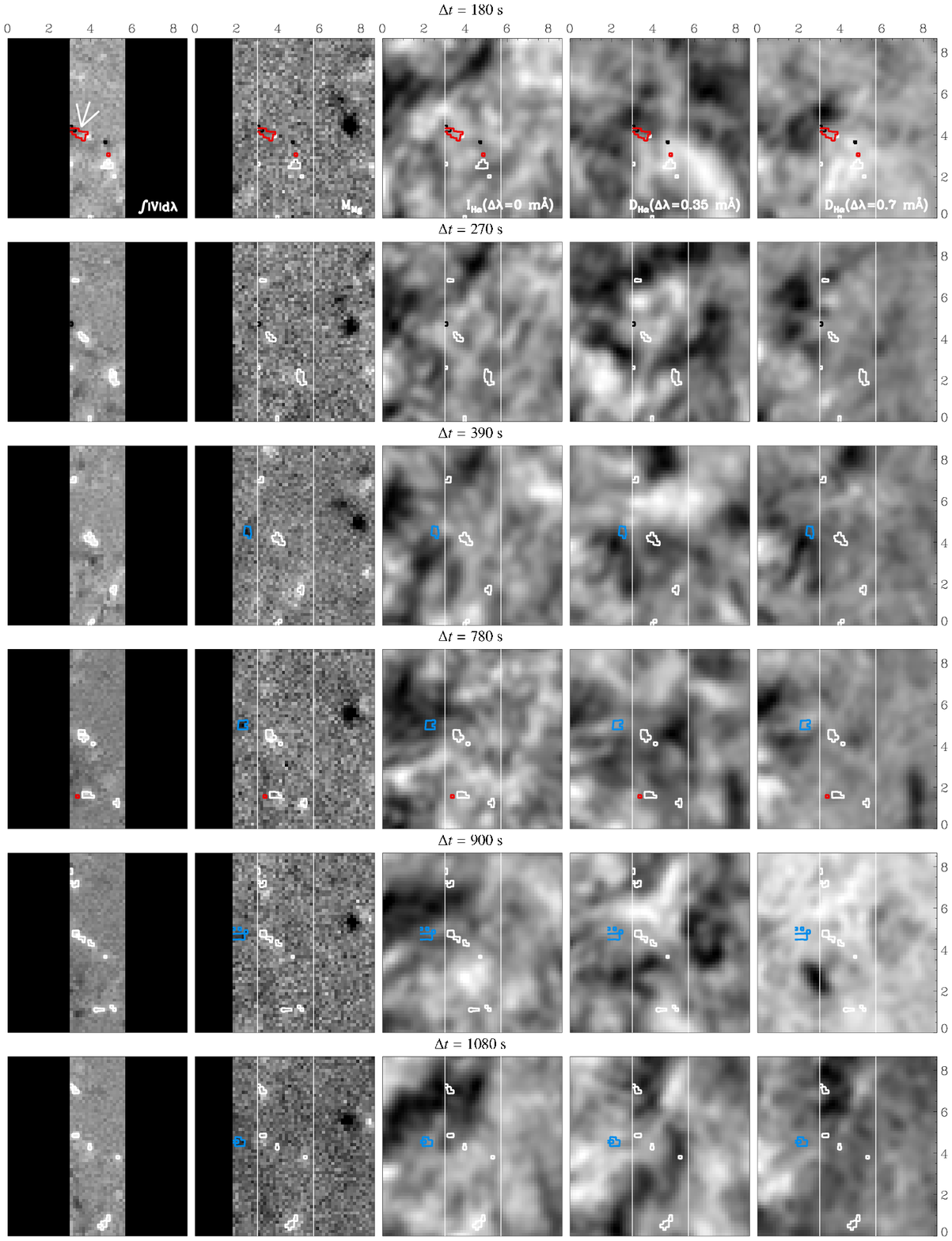}
\caption{Simultaneous observations of the magnetic loop ml20 emerging in the quiet Sun by {\it Hinode}
and the Dutch Open Telescope. The loop appeared on September 26, 2007 at 09:06 UT. From left
to right: circular polarization map in \ion{Fe}{1} 630.25~nm, magnetic flux density from the
\ion{Mg}{1}~b magnetograms (scaled to $\pm 22.7$ Mx cm$^{-2}$), H$\alpha$ line core image, and 
H$\alpha$ Dopplergrams at $\Delta \lambda=\pm 0.35$ and $\Delta \lambda=\pm 0.7$ \AA\/ from 
line center, respectively.} 
\label{ej2_crom_todos}
\end{figure*}

\section{Effects in the upper chromosphere}
In the previous section we have seen that small-scale magnetic loops
in the quiet Sun may rise through the atmosphere and reach the layers
where the central part of the \ion{Mg}{1}~b 517.3~nm line is
formed. Some of them also produce \ion{Ca}{2}~H brightness
enhancements. In this context, the question naturally arises as to the
maximum height that these structures can attain. Are they able to
reach the upper chromosphere or even the corona?

Here we use the H$\alpha$ observations of the DOT to provide a partial
answer to this question. Unfortunately, there is little overlap
between the {\it Hinode} and DOT measurements because of bad seeing
conditions. Nevertheless, for one of the loops detected by {\it
Hinode} there is simultaneous coverage from the DOT. We use these data
to attempt to observe the rise of the loop to the upper chromosphere. 
The analysis is not complete and must be refined with EUV and X-ray 
observations tailored to the detection of such magnetic
structures in the hot corona.

The loop observed simultaneously by {\it Hinode} and the DOT (ml20) appeared on September
26, 2007, at 09:06 UT. In Fig.\ \ref{ej2_crom_todos} we show its
evolution in the photosphere/temperature minimum region using the
\ion{Fe}{1} 630.25~nm circular polarization maps and the \ion{Mg}{1}~b 
magnetograms (first and second columns, respectively). The figure also
displays H$\alpha$ line core filtergrams, as well as H$\alpha$
Dopplergrams at $\pm 0.35$ \AA\/ and $\pm 0.7$ \AA\/ from line
center. If the loop reaches the layers where H$\alpha$ is formed, it
should first appear in the fifth column, then in the fourth, and
finally in the third.

The loop emerged as a small patch of linear polarization at the border
of a granule ($\Delta t = 0$ s, not shown). Its subsequent evolution
is similar to that of the loop considered in Fig.\ \ref{ej1_crom}. The
footpoints of the loop are detected in the photosphere for the first
time at $\Delta t= 180$ s. The whole structure shows upflows in the
Stokes V zero-crossing maps, indicating its ascent. Unfortunately, the
negative footpoint is close to the border of the FOV scanned by the SP
and soon disappears from the photospheric maps.  The loop becomes
visible in the \ion{Mg}{1}~b magnetograms at $\Delta t = 390$~s. At
this time there is clear signal in the negative footpoint and weaker
polarization in the positive leg. At $\Delta t=780$~s both footpoints
show stronger magnetogram signals but the distance between them has
not increased. The ascent of the loop to the 
chromosphere is associated with downflows in the \ion{Mg}{1}~b Dopplergram and 
brightenings in the Ca\,{\sc ii}\,{\sc H} line-core images.
The \ion{Fe}{1} and \ion{Mg}{1}~b
signals start to fade at $\Delta t = 900$ s until the loop disappears
simultaneously from the low and the upper photospheric layers.

The H$\alpha$ line core images
and the Dopplergrams do not show any particular feature that could be
associated with the appearence of an arch filament system in the
chromosphere. It is important to remark that the maximum separation
between the footpoints of this loop was only 760 km. It may well be
that a larger separation is required for the apex of the loop to reach
the upper chromosphere. In fact, excessive magnetic tension might
prevent the field lines from rising. Keeping in mind these
considerations, we do not discard that magnetic loops with larger
separations may be seen in future H$\alpha$ observations.

\section{Physical properties of the emerging loops}

In this section we characterize the physical properties of the
small-scale magnetic loops observed with {\it Hinode}. Weak exam\-ples
and loops appearing close to strong network elements or in crowded
areas are omitted from the analysis to maintain the quality of the
results. This leaves us with 33 loops, which represents 48\% of the
total sample.

Table \ref{tabla} lists the basic parameters of the loops, including
lifetimes ($\Delta t$), maximum distances between footpoints ($D_{\rm
max}$), speeds at which the footpoints separate initially ($V_{\rm
D}$), total magnetic fluxes ($\Phi_{\rm 630}$) and maximum flux
densities ($\phi_{\rm 630}$) in the photosphere, estimations of maximum magnetic flux
densities at the height of formation of the \ion{Mg}{1}~b measurements
($\phi_{\rm Mg}$), and an estimation of the largest downflows detected in the
\ion{Mg}{1}~b Dopplergrams ($v_{\rm Mg}$). The lifetime is the time
elapsed between the appearance and disappearance of the polarization
signals. When two numbers are given, the first indicates the time
passed until one of the footpoints interacts with a neighboring
magnetic element. The second is the time of disappearance proper; if
it is accompanied by an asterisk, then the polarization signatures of
the loop were still visible at the end of the observations. To compute
the total magnetic flux we define the footpoints as those regions
where the flux density is larger than $3 \, \sigma_\phi$ at the
position of the loop. The value of $\Phi_{630}$ reported in Table
\ref{tabla} is the maximum flux detected in one of the footpoints 
during the loop evolution, and the error indicates the uncertainty in
$\Phi_{630}$ caused by photon noise.

The last four columns of Table \ref{tabla} give the time intervals
between the appearance of the loops in the photosphere and their
detection in the \ion{Mg}{1}~b magnetograms, the \ion{Mg}{1}~b
Dopplergrams , the \ion{Ca}{2}~H line core images, and the CN filtergrams. 
We consider that a loop is present in any of these 
maps when at least one of the footpoints shows up clearly. The two 
numbers in each column correspond to the positive footpoint (left) 
and the negative one (right).

\renewcommand{\tabcolsep}{.38em}

\begin{table*}[!t]
\begin{minipage}{\linewidth}
\begin{center}
\caption{Physical properties of the small-scale emerging magnetic loops}
\label{tabla}
\begin{tabular}{cccccccccccccc}
\tableline
Name & Date & t$_\mathrm{ini}$ & $\Delta$t & D$_\mathrm{max}$ & $V_\mathrm{D}$ & $\Phi_\mathrm{630}$ & $\phi_\mathrm{630}$ & $\phi_\mathrm{Mg}$ & $V_{\rm Mg}$ & $\Delta$t$_\mathrm{SP-Mg}$ & $\Delta$t$_\mathrm{SP-down}$ & $\Delta$t$_\mathrm{SP-Ca}$ & $\Delta$t$_\mathrm{SP-CN}$  \\
 & (09/07) & (UT) & (s) & (km) & (km/s)  & (Mx) & (Mx/cm$^2$) & (Mx/cm$^2$) & (km/s) & (s) & (s) & (s) & (s) \\
\tableline
ml01 & 25 & 13:36:15 & 240       & 800  & 3.9 &  4.6$\times$10$^{16}\pm 1\times$10$^{15}$ & $23.2\pm 0.6$ &              &      &         &         &        &             \\
ml02 & 25 & 13:35:45 & 150/1020  & 1220 & 4.0 &  1.5$\times$10$^{17}\pm 2\times$10$^{15}$ & $24.8\pm 0.6$ &              &      &         &         &        &          \\
ml03 & 25 & 13:37:45 & 330       & 560  & 0.9 &  7.6$\times$10$^{16}\pm 1\times$10$^{15}$ & $29.7\pm 0.4$ &              &      &         &         &        &             \\
ml04 & 25 & 13:46:15 & 60/480    & 690  & 1.7 &  8.2$\times$10$^{16}\pm 1\times$10$^{15}$ & $23.1\pm 0.9$ &              &      &         &         &        &             \\
ml05 & 25 & 13:45:15 & 240       & 790  & 1.1 &  1.1$\times$10$^{17}\pm 1\times$10$^{15}$ & $23.0\pm 1.1$ &              &      &         &         &        &             \\
ml06 & 25 & 13:36:15 & 180       & 910  & 6.0 &  3.6$\times$10$^{16}\pm 9\times$10$^{14}$ & $16.6\pm 0.6$ &              &      &         &         &        &              \\
ml07 & 25 & 13:54:15 & 90        & 490  & 1.4 &  3.5$\times$10$^{16}\pm 8\times$10$^{14}$ & $8.1\pm 0.3$  &              &      &         &         &        &              \\

ml08 & 25 & 13:41:45 & 630       & 800  & 6.2 &  3.9$\times$10$^{16}\pm 1\times$10$^{15}$ & $25.3\pm 0.6$ &              &      &         &         &        &               \\

ml09 & 25 & 14:39:30 & 960/1200  & 4000 & 3.9 &  1.3$\times$10$^{17}\pm 1\times$10$^{15}$ & $32.8\pm 0.4$ & 15.0$\pm1.1$ &-1.1  & 270/540 & 270/750 &780/780 & 780/1230  \\
ml10 & 25 & 14:42:00 & 120       & 560  & 1.7 &  1.5$\times$10$^{16}\pm 6\times$10$^{14}$ & $18.6\pm 1.1$ &              &      &         &         &        &          \\
ml11 & 25 & 14:51:00 & 300/2550* & 1670 & 2.5 &  1.2$\times$10$^{17}\pm 1\times$10$^{15}$ & $32.3\pm 0.6$ & 10.4$\pm1.0$ &-0.58 & 180/450 & 180/450 &420/660 & /660     \\
ml12 & 25 & 14:51:00 & 300/2550* & 2040 & 0.7 & 1.3$\times$10$^{17}\pm 1\times$10$^{15}$  & $24.3\pm 0.4$ & 8.3$\pm$0.8  &      & 690/750 &         &        &          \\
ml13 & 25 & 15:04:00 & 660/1230  & 1440 & 2.1 &  7.8$\times$10$^{16}\pm 1\times$10$^{15}$ & $28.9\pm 0.5$ & 28.6$\pm$0.9 &-0.70 & 390/390 & 690/    &750/    & 690/      \\
ml14 & 25 & 15:06:00 & 180       & 480  & 3.7 &  6.1$\times$10$^{16}\pm 1\times$10$^{15}$ & $23.4\pm 0.4$ &              &      &         &         &        &          \\
ml15 & 25 & 15:14:00 & 360/1590* & 3170 & 0.5 &  2.0$\times$10$^{17}\pm 2\times$10$^{15}$ & $29.8\pm 0.4$ & 5.4$\pm$0.5  &-0.91 & 330/330 & 570/420 &750/420 & 690/570   \\
ml16 & 25 & 15:24:30 & 510       & 790  & 0.9 &  1.3$\times$10$^{17}\pm 2\times$10$^{15}$ & $34.2\pm 0.4$ & 7.3$\pm$0.5  &-0.59 & 240/240 & 270/270 &        &          \\
ml17 & 25 & 14:50:30 & 90        & 620  & 1.0 &  5.5$\times$10$^{16}\pm 1\times$10$^{15}$ & $22.6\pm 0.5$ &              &      &         &         &        &          \\
                                                                                                                                    
ml18 & 26 & 08:32:00 & 510/1110* & 990  & 1.1&   1.1$\times$10$^{17}\pm 1\times$10$^{15}$ & $33.5\pm 0.4$ &              &      &         &         &        &          \\

ml19 & 26 & 08:23:30 & 660       & 1670 & 3.1 &  6.2$\times$10$^{16}\pm 1\times$10$^{15}$ & $18.1\pm 0.4$ & 10.0$\pm$0.9 &-0.33 & /180    & /180    &/210    & /180     \\

ml20 & 26 & 09:06:00 & 1050      & 1450 & 2.2 &  1.5$\times$10$^{16}\pm 6\times$10$^{14}$ & $15.7\pm 0.9$ & 5.0$\pm$0.4  &-0.38 & 390/360 & 510/390 &570/420 & 600/390   \\

ml21 & 26 & 09:47:30 & 240       & 700  & 2.8 &  3.8$\times$10$^{16}\pm 9\times$10$^{14}$ & $23.7\pm 0.6$ &              &      &         &         &        &          \\

ml22 & 26 & 11:24:00 & 1380*     & 1570 & 0.9 &  2.4$\times$10$^{17}\pm 2\times$10$^{15}$ & $42.2\pm 0.3$ & 11.8$\pm$1.1 &-0.48 & 360/450 & 600/750 &630/750 &           \\

ml23 & 26 & 11:48:00 & 150       & 520  & 3.6 &  3.3$\times$10$^{16}\pm 9\times$10$^{14}$ & $23.1\pm 0.6$ &              &      &         &         &        &          \\

ml24 & 26 & 12:16:00 & 120/390   & 830  & 3.4 &  1.9$\times$10$^{18}\pm 2\times$10$^{15}$ & $75.8\pm 1.1$ &              &      &         &         &        &          \\

ml25 & 26 & 12:28:00 & 1230      & 2840 & 0.1 &  2.0$\times$10$^{16}\pm 7\times$10$^{14}$ & $21.4\pm 0.8$ & 7.50$\pm$0.8 &-0.56 & 210/420 & 540/630 &690/690 &          \\

ml26 & 27 & 09:25:00 & 1110      & 760  & 0.9 &  7.0$\times$10$^{16}\pm 1\times$10$^{15}$ & $32.7\pm 0.6$ & 26.7$\pm$0.8 &-0.42 & 270/270 & 540/    &        &          \\

ml27 & 27 & 11:08:00 & 420       & 960  & 4.5 &  8.6$\times$10$^{16}\pm 1\times$10$^{15}$ & $30.6\pm 0.6$ &              &      &         &         &        &          \\

ml28 & 27 & 12:20:30 & 90        & 670  & 0.9 &  3.0$\times$10$^{16}\pm 9\times$10$^{14}$ & $15.2\pm 0.5$ &              &      &         &         &        &          \\

ml29 & 27 & 12:26:30 & 1110      & 2290 & 4.0 &  1.7$\times$10$^{17}\pm 2\times$10$^{15}$ & $27.2\pm 0.4$ & 14.5$\pm$0.8 &-0.49 & 150/150 & 450/390 &/450    &          \\

ml30 & 28 & 11:28:30 & 300/750   & 1300 & 0.9 &  7.8$\times$10$^{16}\pm 1\times$10$^{15}$ & $24.4\pm 0.5$ &              &      &         &         &        &          \\
ml31 & 28 & 11:38:00 & 180       & 1560 & 1.5 &  7.8$\times$10$^{16}\pm 1\times$10$^{15}$ & $19.0\pm 0.3$ &              &      &         &         &        &          \\

ml32 & 28 & 12:32:30 & 720/720   & 710  & 1.5 &  1.9$\times$10$^{17}\pm 1\times$10$^{15}$ & $44.6\pm 0.4$ & 19.0$\pm$0.5 &      & 240/240 &         &360/    & 330/     \\

ml33 & 29 & 08:39:00 & 240/630   & 860  & 2.4 &  2.1$\times$10$^{17}\pm 1\times$10$^{15}$ & $42.2\pm 0.3$ &              &      &         &         &        &          \\ 
\tableline

{\bf Mean} &   &       & {\bf 741}   & {\bf 1234}     & {\bf 2.2} &  {\bf 9.13$\times$10$^{16}$} & {\bf 26.1} & {\bf 13.0} & {\bf -0.60} & {\bf 295} & {\bf 406} & {\bf 513} &{\bf 514}    \\ 
\tableline\\
\end{tabular}
\end{center}
\end{minipage} 
\end{table*}

As can be seen in Table \ref{tabla}, there is a wide range of loop
parameters. The lifetimes vary from some 2 min up to 40 min, although
most of the loops disappear in less than 10 minutes. The maximum
separation between the footpoints is a strong function of the lifetime
and ranges from $\sim 500$~km to 4000 km. Many loops reach horizontal
dimensions comparable to, or larger than, those of granules. Therefore, they
must be viewed as coherent structures capable of withstanding the
conditions of the granular environment for a relatively long time. The
initial velocity of separation between the footpoints does not seem to
have any relationship with the other parameters listed in the
table. Values of 0.1 to 6~km~s$^{-1}$ are typical. As already mentioned,  
the separation speed tends to decrease when the footpoints reach the
intergranular space, likely because horizontal motions there are not
as vigorous as in the interior of granular cells.

The longitudinal magnetic fluxes measured in the footpoints range from $2 \times 10^{16}$ to $2 \times 10^{17}$ Mx at the level of formation of the \ion{Fe}{1} 630 nm lines\footnote{Note that the magnetic flux of ml24 is $1.9 \times 10^{18}$ Mx, considerably larger than the rest of values. This might be an artifact caused by the difficult separation of ml24 and the strong network element with which it interacts.}. Therefore, the loops have smaller fluxes than ephemeral
regions and should be placed at the lower end of the flux distribution
observed in emerging active regions. The magnetic flux density in the
footpoints is typically 20--40~Mx~cm$^{-1}$. To infer the magnetic field strength from the magnetic 
flux density values we need to know the filling factor of the 
field lines that build the loop structure and their inclination. The footpoints should be
relatively vertical because of geometrical reasons. Assuming that the fields 
occupy most of the resolution element, i.e., that the magnetic
filling factor is close to unity the field strength of the loops can be estimated to be of
order 10-100~G. Only if the filling factor is much smaller than unity
would the field strength increase to kG values, but we consider this
possibility unlikely in view of Fig.~\ref{hist_ratio_flux}. 

An important result is that 23\% of the loops are detected in the 
\ion{Mg}{1} b magnetograms after their appearance in the photosphere 
(16 cases out of 69). It takes an average of 5 minutes
for the loops to move from the photosphere to the height at which the \ion{Mg}{1} b measurements form, although faster and slower ascents have been observed too. All the loops detected in the \ion{Mg}{1}~b
magnetograms develop downflows at the same heights. In addition, 15\%
of the loops are seen as bright points in \ion{Ca}{2}~H line-core
filtergrams. This means that an important fraction of the magnetic
flux that emerges into the photosphere reaches the chromosphere. As
they travel upward, the loops are observed in the \ion{Fe}{1}
magnetograms, the \ion{Mg}{1}~b magnetograms, the \ion{Mg}{1}~b
Dopplergrams, the \ion{Ca}{2}~H line-core images, and the 
CN filtergrams (in this order).

By contrast, 77\% of the loops never make it to the chromosphere. We
have been unable to identify any parameter determining whether a given
loop will rise or not. This includes the total magnetic flux and the
magnetic flux density. However, low-lying loops tend to have lifetimes
shorter than 500~s and separations smaller than 500 km. Thus it may
simply be that they do not last long enough to reach high atmospheric
layers. Further work is clearly needed to explain why a substantial
fraction of the loops remain low-lying. Also, the relation between
these structures and the transient horizontal fields described by
\cite{ishikawa+tsuneta2009} should be investigated, given their
similar lifetimes and magnetic topologies.

\begin{figure}[!t]
\centering{
\includegraphics[width=1.2\columnwidth]{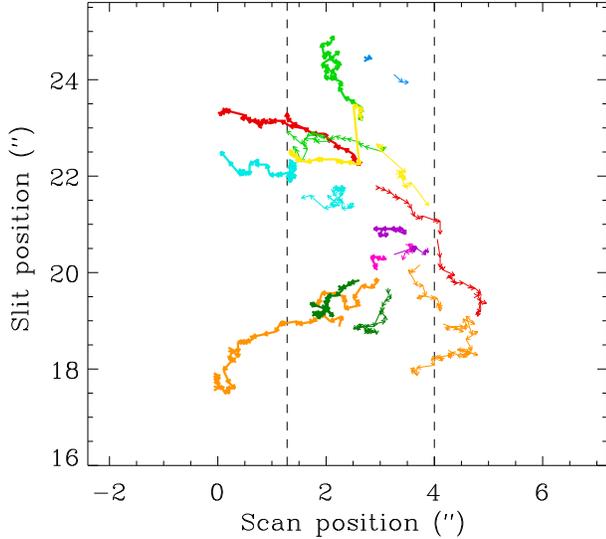}}
\caption{Trajectories of 9 loops that appeared on September 25, 2007 (see the Animation in 
the electronic edition of the Astrophysical Journal). Thick and thin lines indicating the positive
and negative footpoints of each loop. The paths in red, blue, light green, yellow, turquoise, pink, orange, dark green and violet correspond to loops ml09 to ml17 in Table \ref{tabla}, respectively. The vertical dashed lines mark
the FOV of the SP observations at 630 nm.}
\label{trayectorias}
\end{figure}

\section{Sites of emergence, evolution, and tilt angles}

The observations described above demonstrate that magnetic fields do
emerge into the quiet solar atmosphere in the form of small-scale
loops, confirming the results of \cite{marian_07} and
\cite{rebe_07}. The loops are detected as a patch of linear
polarization flanked by two circular polarization signals of opposite
polarity. In nearly all the cases the linear polarization appears
before or at the same time than the Stokes $V$ signals, as can be
expected from $\Omega$-shaped loops rising through the
atmosphere. Only in two cases out of 69 have we detected linear
polarization after the loop had already disappeared. In those cases,
the footpoints were approaching each other. This behavior is
compatible with a loop that emerges and then submerges in the
photosphere, or with a ``magnetic bubble'', i.e., a circle of magnetic
field lines.

The long duration of our time series has permitted us to discover the
existence of emergence centers in which several loops appear one after
the other. For example, there is a $8\arcsec \times 3\arcsec$ region
of the solar surface where we have detected 9 events in a time
interval of 1 h. The complete set of observations covering this period
is provided as an mpeg animation in the electronic edition of the
Astrophysical Journal. Some of the loops even appear at the very same
position. The example shown in Fig.\ \ref{ej1_crom} belongs to this
area. The existence of emergence centers may have important
consequences for the origin of the loops. These regions act as 
subsurface reservoirs of magnetic flux that is transferred 
intermittently to the photosphere by an as yet
unkwnown mechanism.

The loops generally emerge in granules or at their edges, although
there are exceptions of loops appearing in dark areas. As the loops
emerge the footpoints separate and the linear polarization fades
away. In most cases, the footpoints do not describe rectilinear
trajectories. Figure \ref{trayectorias} shows the paths followed by
the two polarities of the loops observed on September 25, 2007 (loops ml09 to ml17 in the Table 2). 
The red curves correspond to the example discussed in Fig.\
\ref{ej1_crom}. In this case the footpoints described quite a
rectilinear path, similarly to emerging active regions and ephemeral
regions. However, the majority of loops show more complicated
trajectories. The reason is that, in general, they emerge in granules
and drift toward the closest intergranular lane. When the footpoints
reach the intergranular space they stay there and are passively
advected by the flow. This creates complicated trajectories. The
important point, however, is that the magnetic field is sufficiently
weak as to be pushed and moved around by the granular flow, but {\it
without being destroyed in the process}. The loops remain coherent
during all their lifetime, as if the granular flow did not exist.

During their evolution, the loops interact with other magnetic flux
concentrations that cross their paths. If the loops stay long in the
photosphere, the footpoints cancel with elements of opposite polarity
or are absorbed by patches of the same polarity. Loops that experience
a fast evolution have more probabilities of avoiding other magnetic
elements and often disappear without undergoing any interaction.

\begin{figure}[!t]
\centering{
\includegraphics[width=\columnwidth,bb=39 12 477 352]{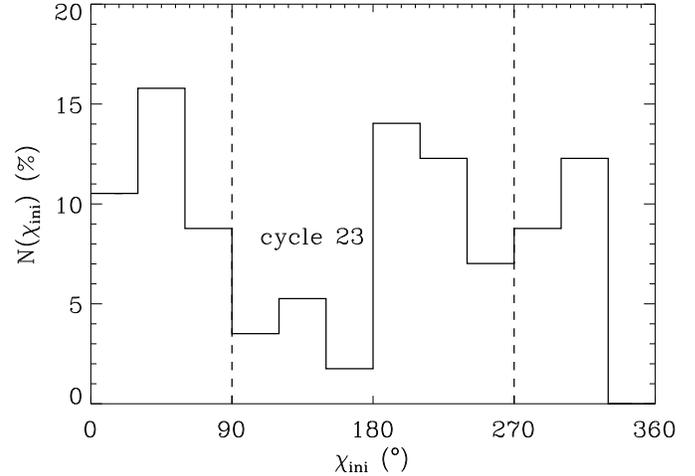}}
\caption{Distribution of initial tilt angles for 66 emerging loops. Angles 
between 90$^\circ$ and 270$^\circ$ imply that the positive polarity patch is the leading one
(in the sense of rotation), consistent with Hale's rules in the
northern hemisphere during solar cycle 23.}
\label{tilt}
\end{figure}

In seeking the origin of the loops it is of interest to determine the
magnetic orientation of their footpoints. If the loops are caused by
the global solar dynamo, one may expect a regular ordering of the
footpoints at the moment of emergence. This is what happens in 
active regions, where the signs of the leader and follower polarities 
are governed by Hale's rules \citep[see, e.g.,][]{stix}.  All the loops
considered here appeared in the northern hemisphere during solar cycle
23. In that cycle, leader polarities were positive in the northern
hemisphere and negative in the southern hemisphere. Figure \ref{tilt}
shows an histogram of the orientation of our small-scale emerging
loops. The tilt angle is defined to be the angle between the solar
equator and the line joining the positive footpoint with the negative
one, measured from the west. The angles compatible with the
orientation of sunspot polarities during solar cycle 23 in the
northern hemisphere are those between 90 and 270$^\circ$. Even if the
statistical sample is not very large, the loops seem to have
nearly random orientations. Thus, we conclude that they do not obey
Hale's polarity rules, much in the same way as the shortest-lived
ephemeral regions \citep{martin_79, hagenaar_03}.

\section{Discussion and conclusions}
Recent observations of Hanle-sensitive lines \citep{javier_04} and
Zeeman-sensitive lines \citep{marian_08, david_07, litesetal_08,
ishikawa+tsuneta2009} suggest that a significant fraction of the quiet
Sun is occupied by magnetic fields. Apparently, these fields are weak
and isotropically distributed in inclination \citep{marian_andres_08}. One way to
shed light on their nature is to study how they emerge in the surface
and what their contribution is to the energy budget of the solar
atmosphere.

On granular scales, magnetic flux appears in the solar photosphere as
transient horizontal fields \citep{ishikawa+tsuneta2009} and
small-scale magnetic loops \citep{marian_07, rebe_07}. We have studied the latter
in detail using seeing-free observations made by {\it Hinode}. In 28
hours of {\it Hinode} data we have detected 69 small-scale loops
emerging in a quiet Sun region of size $2.7\arcsec
\times 40.6\arcsec$ at disk center. The occurrence rate is 
thus 0.02 events hr$^{-1}$ arcsec$^{-2}$. The loops show clear
spectropolarimetric signatures with a central region of linear
polarization and two patches of circular polarization of opposite
polarity. 

The longitudinal flux observed in each footpoint ranges from $2 \times 10^{16}$ to $2 \times 10^{17}$ Mx, 
with an average of $\sim 9.1 \times 10^{16}$ Mx. This means that the loops represent the smallest emerging flux
regions detected to date (ephemeral regions have fluxes above
10$^{18}$ Mx; Zwaan 1987). The rate at which magnetic flux is carried
to the quiet photosphere by the loops can be estimated to be $1.1
\times 10^{12}$~Mx~s$^{-1}$~arcsec$^{-2}$, or $1.1 \times 10^{24}$~Mx
over the solar surface per day. This is about half the value derived by Lites et al. (1996) for 
horizontal internetwork fields, but still enormous (see \citet{vandriel_02} for a comparison with the flux emergence rates in active and ephemeral
regions).

In the photosphere, the linear polarization associated with the top of
the loop disappears soon, while the circular signals tracing the loop
legs are observed to separate with time. This behavior is consistent
with field lines moving upward through the solar atmosphere. Also the
upflows observed in the Stokes V zero-crossing velocities at the
position of the footpoints confirm the ascent of the loops. 23\% of the loops are detected in \ion{Mg}{1}~b magnetograms
that sample the upper photosphere or the temperature minimum region
(say, 400~km above the continuum forming layer). There is a time delay
of about 5 minutes between the first detection in the photosphere and
the appearance in the \ion{Mg}{1}~b magnetograms, implying an ascent
speed of the order of 1~km~$^{-1}$. Some of the loops continue to
travel upward and become visible in \ion{Ca}{2}~H line-core
filtergrams as small brightness enhancements. Thus, a fraction of 
the loops are able to reach the low chromosphere, carrying 
magnetic flux with them.

The rise of small-scale magnetic loops may provide an efficient
mechanism to transfer substantial amounts of energy from the
photosphere to the chromosphere. This would support claims by
\cite{javier_04} and \cite{ishikawa+tsuneta2009} that the tangled fields of the quiet Sun store
sufficient energy to heat the chromosphere. A related question is whether the small-scale loops
rise up to the transition region or even the corona. The observations
required to answer this question are quite challenging due to the
different spatial resolutions attainable with present day
optical, EUV, and X-ray instruments, but should be pursued.

About 77\% of the loops that appear in the solar surface never rise to
the chromosphere. These loops have the shortest lifetimes and show the
smallest footpoint separations; other than that, they do not differ
from those reaching higher layers. Usually, they disappear close to
their emergence sites. The fields associated with these loops might
represent the tangled quiet Sun fields deduced from Hanle
measurements \citep{stenflo_87, rafa_04}, but a definite
conclusion cannot be made without studying the compatibility of Hanle and 
Zeeman measurements.

What is the origin of the small-scale magnetic loops? One possibility
is that they are created by the solar dynamo at the bottom of the
convection zone, as part of a larger toroidal flux
tube. \cite{mark_07} presented three-dimensional MHD simulations of the
last stages of the emergence of one such tube. They placed a
horizontal tube at the top of the convection zone, just beneath the
photosphere. When the initial magnetic flux is smaller than
$\sim$10$^{18}$ Mx, the tube is not sufficiently buoyant to rise
coherently against the convective flows and fragments. At the surface,
the process of flux emergence occurs on very small spatial scales
(typically 1000-2000 km) and short time scales (5 min). These
properties are compatible with our observations. Thus, the small-scale
loops we have detected may simply be the result of weak flux tubes
distorted by the granulation as they emerge from the convection zone
into the photosphere. The fragmentation of the tubes might explain why
there are emergence centers where loops appear recurrently one after
the other. A preliminary analysis of the footpoint
orientations suggests that the loops do not show a tendency
to be aligned according to Hale's rules. This can be regarded as a
considerable difficulty against the idea that the origin of the loops
is the solar dynamo. However, it may also be a natural consequence of
the interaction of the tube's fragments with the near-surface 
granular convection if it removes all the information 
carried originally by the tube.

Another possibility is that the magnetic loops represent flux recycled
from decaying active regions. In a sense, the MHD simulations of
Abbett (2007), \cite{isobe_08}, and \cite{steiner_08} model such a
process, because all of them assume an initial magnetic field in
the computational box which could be provided by decaying active
regions. In the simulations, the field evolve and interact with the
granular flows. This interaction creates a significant amount of horizontal fields, 
even if the initial field is purely vertical. Moreover, the simulations show the emergence of magnetic
loops on granular scales. The loops are less coherent than classical
flux tubes and do not connect to deeply rooted field lines. In this
scenario, the magnetic fields of the quiet Sun, and thus the emergence
events we have described, would be the consequence of local
processes acting on the remnants of decaying active regions.

Yet another possibility is that the loops represent submerged
horizontal magnetic fields carried to the surface by the upward
motions of granules or by magnetic buoyancy, as modeled by
\citep{steiner_08}. Even in that case, the origin of such 
submerged fields would be unknown. 

Nowadays, we do not have enough observational
constraints to distinguish between a surface dynamo or a "exploding"
magnetic flux tube emerging from the solar interior. Determining the nature of 
the magnetic loops observed in internetwork regions is important for a better understanding 
of the magnetism of the quiet Sun and its role in the heating of the solar atmosphere. Future efforts should concentrate on the solution of these problems. In addition to high-resolution photospheric observations, polarization measurements in the chromosphere are recquired to track the evolution of the loops with height. These data can now be provided by two-dimensional spectrometers like IBIS, CRISP, or IMaX.

\begin{acknowledgements}

We thank Andr\'es Asensio Ramos, Pascal D\'emoulin and Rafael Manso Sainz for very
helpful discussions, and V\'eronique Bommier for carefully reading the
manuscript. We are grateful to all the observers who participated in
the {\it Hinode} Operation Plan 14, both at ISAS/JAXA and at the
ground-based telescopes. Special thanks are due to Suguru Kamio (NAOJ)
for coordinating the campaign and to Peter S\"utterlin (Utrecht
University) for making the observations at the Dutch Open Telescope
and reducing them. Hinode is a Japanese mission developed and launched
by ISAS/JAXA, with NAOJ as a domestic partner, and NASA and STFC (UK)
as international partners. It is operated by these agencies in
cooperation with ESA and NSC (Norway). Part of this work was carried
out while one of us (MJMG) was a Visiting Scientist at the Instituto
de Astrof\'{\i}sica de Andaluc\'{\i}a. We acknowledge financial
support from the Spanish MICINN through projects ESP2006-13030-C06-02,
PCI2006-A7-0624, and AYA2007-63881, and from Junta de Andaluc\'{\i}a
through project P07-TEP-2687.

\end{acknowledgements}


\end{document}